\documentclass[11pt]{article}
\usepackage{a4}
\usepackage{amsfonts}
\usepackage{amsmath,amssymb}

\usepackage{graphicx}

\makeatother

\def\un#1{\relax\ifmmode\@@underline#1\else
        $\@@underline{\hbox{#1}}$\relax\fi}

\let\du=\du                     

\def\a{\alpha}

\def\f{\phi}

\def\h{\eta}

\def\k{\kappa}
\def\l{\lambda}
\def\m{\mu}
\def\n{\nu}

\def\p{\pi}

\def\r{\rho}

\def\D{\Delta}

\def\L{\Lambda}

\def\ve{\varepsilon}

\def\bo{{\raise-.3ex\hbox{\large$\Box$}}}               
\def\pa{\partial}                                       
\def\TH{{\raise.2ex\hbox{$\displaystyle \bigodot$}\mskip-4.7mu \llap H \;}}
\def\face{{\raise.2ex\hbox{$\displaystyle \bigodot$}\mskip-2.2mu \llap {$\ddot
        \smile$}}}                                      

\def\abs#1{\left| #1\right|}                    
\def\leftrightarrowfill{$\mathsurround=0pt \mathord\leftarrow \mkern-6mu
        \cleaders\hbox{$\mkern-2mu \mathord- \mkern-2mu$}\hfill
        \mkern-6mu \mathord\rightarrow$}
\def\dvec#1{\vbox{\ialign{##\crcr
        \leftrightarrowfill\crcr\noalign{\kern-1pt\nointerlineskip}
        $\hfil\displaystyle{#1}\hfil$\crcr}}}           
\def\dt#1{{\buildrel {\hbox{\LARGE .}} \over {#1}}}     

\def\frac#1#2{{\textstyle{#1\over\vphantom2\smash{\raise.20ex
        \hbox{$\scriptstyle{#2}$}}}}}                   
\def\sfrac#1#2{{\vphantom1\smash{\lower.5ex\hbox{\small$#1$}}\over
        \vphantom1\smash{\raise.4ex\hbox{\small$#2$}}}} 
\def\bfrac#1#2{{\vphantom1\smash{\lower.5ex\hbox{$#1$}}\over
        \vphantom1\smash{\raise.3ex\hbox{$#2$}}}}       
\def\afrac#1#2{{\vphantom1\smash{\lower.5ex\hbox{$#1$}}\over#2}}    

\def\[{\lfloor{\hskip 0.35pt}\!\!\!\lceil}
\def\]{\rfloor{\hskip 0.35pt}\!\!\!\rceil}

\def\du#1#2{_{#1}{}^{#2}}

\def\fracm#1#2{\hbox{\large{${\frac{{#1}}{{#2}}}$}}}
\def\ha{{\fracmm12}}

\def\un{\underline}
\def\fracmm#1#2{{{#1}\over{#2}}}

\def\low#1{{\raise -3pt\hbox{${\hskip 0.75pt}\!_{#1}$}}}

\def\Dot#1{\buildrel{_{_{\hskip 0.01in}\bullet}}\over{#1}}
\def\dt#1{\Dot{#1}}

\def\DDot#1{\buildrel{_{_{\hskip 0.01in}\bullet\bullet}}\over{#1}}
\def\ddt#1{\DDot{#1}}

\newskip\humongous \humongous=0pt plus 1000pt minus 1000pt

\newif\ifdtup

\newcommand{\be}{\begin{equation}}
\newcommand{\ee}{\end{equation}}
\newcommand{\nbe}{\begin{equation*}}
\newcommand{\nee}{\end{equation*}}

\newcommand{\lb}{\label}

\begin{document}

\thispagestyle{empty}

{\hbox to\hsize{
\vbox{\noindent June 2010 \hfill version 5 }}}

\noindent
\vskip2.0cm
\begin{center}

{\large\bf FOURTH-ORDER GRAVITY AS THE \vglue.1in
INFLATIONARY MODEL REVISITED~\footnote{
Supported in part by the Japanese Society for Promotion of Science (JSPS)}}
\vglue.3in

Sho Kaneda${}^a$, Sergei V. Ketov~${}^{a,b}$, Natsuki Watanabe${}^a$ 
\vglue.1in

${}^a$ {\it Department of Physics, Tokyo Metropolitan University, Japan}\\
${}^b$ {\it IPMU, University of Tokyo, Japan}
\vglue.1in
kaneda-sho@ed.tmu.ac.jp, ketov@tmu.ac.jp, watanabe-natsuki1@ed.tmu.ac.jp
\end{center}

\vglue.3in

\begin{center}
{\Large\bf Abstract}
\end{center}
\vglue.1in

\noindent We revisit the old (fourth-order or quadratically generated) gravity
 model of Starobinsky in four space-time dimensions, and derive the (inflaton) 
scalar potential in the equivalent scalar-tensor gravity model. 
The inflaton scalar potential is used to compute the (CMB) observables of 
inflation, associated with curvature perturbations (namely, the scalar and 
tensor spectral indices, and the tensor-to-scalar ratio), including the new 
{\it next-to-leading-order} terms with respect to the inverse number 
of e-foldings.  The results are compared to the recent (WMAP5) experimental 
bounds. We confirm both mathematical {\it and} physical equivalence between 
$f(R)$ gravity theories and the corresponding scalar-tensor gravity theories.

\newpage

\section{Introduction}

{\it Inflation} is a proposal (cosmological paradigm) about the existence of 
a short but fast (exponential, or de-Sitter-type) accelerated grow of the
FLRW scale factor $a(t)$ in the early Universe, after the Big-Bang but before 
the radiation-dominated epoch \cite{inf}. It implies  
\be \lb{definf} \ddt{a}(t)>0 \ee
Though the whole idea of inflation remains to be a speculation, there is the 
significant (indirect) evidence for it. In the first place, it is the correct 
prediction of CMB fluctuations and large scale structure, in remarkable 
agreement with the WMAP observations of CMB --- see eg., ref.~\cite{wmap}. 
Inflation can generate irregularities in the Universe that may lead to the 
formation of structure.  The main discriminators among various inflationary 
models are the {\it spectral indices} associated with the primordial power 
spectrum of curvature perturbations \cite{llbook}. For instance, the on-going 
PLANCK satellite mission is going to provide tight constraints on the 
observable spectral indices with the accuracy of under $0.5$ percent 
\cite{nature}. Though the basic formulae for the spectral indices in terms
of any inflaton potential are well known \cite{llbook}, their dependence upon
the e-foldings number can only be computed in a specific inflationary model.
Our motivation here is to reconsider primary candidates among the inflationary 
models, as to whether they can survive precisional tests in a near future. 

The excellent model of chaotic inflation was proposed by Starobinsky in 1980
\cite{star}. It is the simplest version of $f(R)$ gravity theories \cite{sot}, 
whose extra term beyond the standard Einstein-Hilbert term is {\it quadratic}
 in the scalar curvature. The Starobinsky model is reviewed in Sec.~2, where we
 also argue why the other (Ricci- and Riemann- curvature) terms in the 
quadratically generated gravitational action are irrelevant to the FLRW 
dynamics. 

Any $f(R)$ gravity model is known to be {\it mathematically} equivalent to 
the certain scalar-tensor gravity via a Legendre-Weyl transform \cite{oldr}. We
 review that procedure in Sec.~3. However, even in the current literature on
the $f(R)$ gravity (see ref.~\cite{sot} and references therein), its {\it 
physical} equivalence to scalar-tensor gravity is put into doubt. As is known 
in Field Theory, any two field theories, related by a field redefinition
or via duality, have {\it the same} observables. In other words, the field
theories that are mathematically equivalent are also physically equivalent. Of
course, in specific cases the full equivalence may be very tricky (cf., for 
instance, the AdS/CFT correspondence), so it still makes sense to calculate the
observables in both equivalent theories. The spectral indices (in the leading 
approximation) of the Starobinsky model were calculated on the $f(R)$ gravity
side a long time ago \cite{mchi}. In this paper we do a calculation on the 
corresponding scalar-tensor gravity side. We confirm the leading terms found in
ref.~\cite{mchi}, and calculate the sub-leading corrections to them in Sec.~4,
with respect to the inverse number of e-foldings. Checking the physical 
equivalence (ie. the same spectral indices) is yet another motivation to our 
paper.

\newpage

\section{Starobinsky model}

There is {\it a priori} no reason of restricting the gravitational Lagrangian 
to the standard Einstein-Hilbert term that is linear in the scalar curvature, 
as long as it does not contradict an experiment. The first attempt of that
kind was made by Weyl as early as 1921. Nowadays, there is no doubt that the 
extra 
terms of the higher-order in the curvature should appear in the gravitational 
effective action of {\it any\/} Quantum Theory of Gravity. For instance, they 
do appear in String Theory --- see eg., ref.~\cite{ket} for a review. Since the
scale of inflation is just a few orders less than the Planck scale 
\cite{llbook}, it is conceivable that the higher-order gravitational terms 
 may be instrumental for inflation. It is already the case in the simplest
 modified gravity model having only the terms quadratic in the curvature 
\cite{oldr}.

As is well known, there exist only three independent {\it quadratic} curvature 
invariants, $R^{\m\n\l\r}R_{\m\n\l\r}$, $R^{\m\n}R_{\m\n}$ and $R^2$. 
In addition, in four space-time dimensions,
\be \lb{euler}
 \int d^4x\, \sqrt{-g} \left( R^{\m\n\l\r}R_{\m\n\l\r} -4R^{\m\n}R_{\m\n}
 +R^2\right) \ee
is  topological for any metric, whereas
\be \lb{extratop}
 \int d^4x\, \sqrt{-g} \left( 3R^{\m\n}R_{\m\n} -R^2\right) \ee
is topological for any FLRW metric. Those combinations do not contribute to the
 (Friedmann) equation of motion for the scale factor, indicating that the 
scalar curvature models play the most important role in cosmological dynamics.
Hence, the most general gravitational action of the highest order $2$ in the 
curvature, which may be relevant for inflation, is given by
\be \lb{action}
S= \fracmm{1}{2\k^2} \int d^4x\, \sqrt{-g} 
\left( 2\L - R  + \alpha R^2 \right)  \ee
where we have introduced the cosmological constant $\L$ and the dimensional
parameter $\a\equiv M^{-2}$ of mass dimension $(-2)$.  We use the spacetime 
signature $(+,-,-,-)$ and the units $\hbar=c=1$. The Einstein-Hilbert term in 
eq.~(\ref{action}) has the standard normalization with $\k=M_{\rm Pl}^{-1}$ 
in terms of the reduced Planck mass $M_{\rm Pl}^{-2}=8\p G_N$. The rest of our 
notation for space-time (Riemann) geometry is the same as in 
ref.~\cite{landau}.    

The model (\ref{action}) is the simplest representative of the Starobinsky 
models \cite{star}. As was shown in refs.~\cite{star,oldr}, the equations of 
motion for the action (\ref{action}) have an inflationary solution with 
$\a\neq 0$ (even when $\L=0$), which is stable provided that $\a>0$.  The 
stability is confirmed by our method in Sec.~3.

\section{f(R) gravity and inflaton}

The model (\ref{action}) is the simplest particular case of the $f(R)$
 gravity models characterized by an action
\be \lb{fR} S_f = -\fracmm{1}{2\k^2}\int d^4x\, f(R) \ee
with some function $f(R)$ of the scalar curvature. Those models are quite 
popular in the current literature --- see eg., the recent reviews \cite{sot}
and the references therein --- due to their theoretical applications to 
inflation and dark energy.

The gravitational equations of motion derived from the action (\ref{fR}) read
\be \lb{feom}
f'(R)R_{\m\n} -\ha f(R)g_{\m\n}+ g_{\m\n}\bo f'(R) -\nabla_{\m}\nabla_{\n}f'(R)
=0 \ee
where the primes denote differentiation.  Those equations of motion are the 
4th-order differential equations with respect to the metric $g_{\m\n}$ 
(ie. with the higher derivatives). Taking the trace of eq.~(\ref{feom}) yields
\be \lb{ftrace}
\bo f'(R) +\fracm{1}{3}f'(R)R-\fracm{2}{3}f(R) =0 \ee
Hence, in contrast to General Relativity having $f'(R)=const.$, 
in f(R) gravity the field $A=f'(R)$  is {\it dynamical}, ie. it represents 
the independent 
propagating (scalar) degree of freedom. In terms of the fields $(g_{\m\n},A)$ 
the equations of motion are of the 2nd order in the derivatives of the fields. 

In fact, any $f(R)$ gravity is classically (mathematically) equivalent to a  
scalar-tensor gravity \cite{oldr}. The equivalence is established by applying 
a Legendre-Weyl transform. The action (\ref{fR}) is equivalent to \be \lb{rmgr}
S_A = \fracmm{-1}{2\k^2}\int d^4x\,\sqrt{-g}\,\left\{ AR-Z(A)\right\} \ee
where the real scalar $A(x)$ is related to the scalar curvature $R$ by the
Legendre transformation
\be \lb{clt} R=Z'(A) \qquad{\rm and}\qquad f(R)=RA(R)-Z(A(R)) \ee

A Weyl transformation of the metric
\be \lb{weylm}
g_{\m\n}(x)\to \exp \left[ \fracmm{2\k\f(x)}{\sqrt{6}} \right] g_{\m\n}(x)\ee
with the arbitrary field parameter $\f(x)$ yields
\be \lb{weylr}
\sqrt{-g}\,R \to \sqrt{-g}\, \exp \left[ \fracmm{2\k\f(x)}{\sqrt{6}} \right]
\left\{ R -\sqrt{\fracmm{6}{-g}}\pa_{\m}\left(\sqrt{-g}g^{\m\n}\pa_{\n}
\f\right)\k -\k^2g^{\m\n}\pa_{\m}\f\pa_{\n}\f\right\} \ee
Hence, when choosing
\be \lb{ch1}
A(\k\f) = \exp \left[ \fracmm{-2\k\f(x)}{\sqrt{6}} \right]  \ee
and ignoring the total derivative, we can rewrite the action (\ref{rmgr}) 
to the form
\be \lb{stgr}
S_{\f} =  \int d^4x\, \sqrt{-g}\left\{ \fracmm{-R}{2\k^2}
+\fracmm{1}{2}g^{\m\n}\pa_{\m}\f\pa_{\n}\f + \fracmm{1}{2\k^2}
\exp \left[ \fracmm{4\k\f(x)}{\sqrt{6}}\right] Z(A(\k\f)) \right\} \ee
in terms of the physical (and canonically normalized) scalar field $\f(x)$.

Equation (\ref{stgr}) is the standard action of the real dynamical scalar
field $\f(x)$ minimally coupled to Einstein gravity and having the scalar 
potential
\be \lb{poten}
V(\f) = -\fracmm{M^2_{\rm Pl}}{2}\exp \left\{
 \fracmm{4\f}{M_{\rm Pl}\sqrt{6}}\right\}
Z\left( \exp \left[ \fracmm{-2\f}{M_{\rm Pl}\sqrt{6}} \right] \right)
\ee
We are now going to employ it as the scalar-tensor gravity model of inflation. 
In order to explicitly derive the inflaton scalar potential (\ref{poten}), 
one has  to solve for $R$ in terms of $\f$ by inverting the relation
\be \lb{invert}   f'(R) = A(\f) \ee 
that follows from eq.~(\ref{clt}) by differentiation.  In the special case of
\be \lb{ourc}
f(R) = R -2\L -\fracmm{1}{M^2}R^2 \ee
we find 
\be  \lb{spot}  
V(\f) = \left( \fracmm{M^2_{\rm Pl}M^2}{8} +\tilde{\L} \right)
\exp \left\{ \fracmm{2\sqrt{2}\f}{M_{\rm Pl}\sqrt{3}}\right\}
- \fracmm{M^2_{\rm Pl}M^2}{4} 
\exp \left\{ \fracmm{\sqrt{2}\f}{M_{\rm Pl}\sqrt{3}}\right\}
+ \fracmm{M^2_{\rm Pl}M^2}{8}
\ee
where the notation $\tilde{\L}=M^2_{\rm Pl}\L$ has been introduced. 
In terms of the new variable and the parameter,
\be \lb{newvar}
y=\sqrt{\fracmm{2}{3}}\fracmm{\f}{M_{\rm Pl}}\qquad {\rm and}\qquad 
V_0=\fracmm{1}{8}M^2_{\rm Pl}M^2 \ee
respectively, the potential (\ref{spot}) reads
\be \lb{spot2}
v(y)=\fracmm{V(y)}{V_0} = 
\left( 1 +\fracmm{\tilde{\L}}{V_0}\right) e^{2y} - 2e^y +1
\ee 
The scalar potential appears to be bounded from below with the only minimum
at $y=0$ (stability!). It is also sufficiently steep for a slow-roll inflation.
It is the last (third) cosmological term on the right-hand-side of 
eq.~(\ref{spot}) that dominates in the potential during the slow-roll 
inflation (when taken alone,
it gives rise to a de-Sitter inflationary solution), the second term represents
 the 1st-order (leading) correction, and the first term is the 2nd-order 
(subleading) correction.~\footnote{As is clear from eq.~(\ref{spot2}), the 
`initial' cosmological term $\tilde{\L}$ is unimportant during the slow-roll 
inflation. The ratio $\tilde{\L}/V_0$ is also negligible from physical (scale) 
arguments.}  In what follows we ignore $\tilde{\L}$. Then the scalar potential 
for the slow-roll inflation gets simplified to
\be \lb{srpot}
 V(y) = V_0\left( e^{y}-1 \right)^2 
\ee 
A graph of the function $v(y)=e^{2y} - 2e^y + 1$ near its minimum $y=0$ is 
given in Fig.~1. After a shift $\f\to \f+ \f_0$ with  
$2\exp\left[ \sqrt{\fracmm{2}{3}}\fracmm{\f_0}{M_{\rm Pl}}\right]=1$,  
the potential (\ref{srpot}) for the sufficiently negative values of $y$ can be
approximated as
\be \lb{effpot}
V_{\rm eff}(\f) \approx V_0 \left[ 1- \exp\left( \sqrt{\fracmm{2}{3}}
\fracmm{\f}{M_{\rm Pl}} \right) \right] 
\ee
where we have ignored the subleading contribution. The scalar potential 
(\ref{effpot}) is known in the inflationary model building \cite{llbook}. In 
our treatment of Sec.~4 we use the potential (\ref{srpot}).

The $(R+R^2)$ gravity (or Starobinsky) model is known as the excellent model of
 chaotic inflation in early Universe, and its spectral indices in the leading 
approximation are also known \cite{mchi}.~\footnote{Though it is irrelevant 
to the early Universe, the Newtonian (weak field) limit of $f(R)$ gravity and 
that of the corresponding scalar-tensor gravity are also {\it the same}, as can
 be easily verified by the use of eq.~(\ref{invert}).}  In the next Sec.~4 we 
derive those indices in the dual (scalar-tensor gravity) picture, and calculate
 the sub-leading terms.   

\begin{figure}[t]
\centering
\includegraphics[width=5cm,clip]{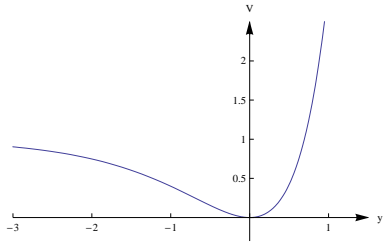}
\caption{\small Graph of the function $v(y)=e^{2y}-2e^y+1$}
\label{fig:1}
\end{figure}

\section{Spectral indices}

The slow-roll inflation parameters are defined by \cite{llbook}
\be \lb{epsi}
\ve(\f) = \fracmm{1}{2} M^2_{\rm Pl} \left( \fracmm{V'}{V}\right)^2
\ee
and
\be \lb{eta}
\h (\f) = M^2_{\rm Pl} \fracmm{V''}{V} 
\ee
where the primes denote the derivatives with respect to the inflaton field 
$\f$. A necessary condition for the slow-roll approximation is the
smallness of the inflation parameters \cite{llbook},
\be \lb{sroll}
 \ve(\f)\ll 1 \qquad {\rm and} \qquad \abs{\h(\f)}\ll 1 \ee
The first condition implies eq.~(\ref{definf}), whereas the second condition
guarantees that inflation lasts long enough, via domination of the friction
term in the inflaton equation of motion (in the slow-roll case):
\be  \lb{fric}
3 H\dt{\f} =- V' \ee
Here $H$ stands for the Hubble `constant' $H(t)=\dt{a}/a$. 
Equation (\ref{fric})
is to be supplemeted by the Friedmann equation 
\be \lb{fried}
 H^2=\fracmm{V}{3M^2_{\rm Pl}}
\ee
It follows from eqs.~(\ref{fric}) and (\ref{fried}) that
\be \lb{feq} 
\dt{\f} =-M_{\rm Pl}\fracmm{V'}{\sqrt{3V}} <0  
\ee
whose solution during the slow-roll inflation $(t_0<t_{\rm start}\leq t \leq
t_{\rm end})$ is
\be \lb{fsol}
\f (t)= -\sqrt{\fracmm{3}{2}}M_{\rm Pl}
\ln \left[ \fracmm{4\sqrt{V_0}}{3\sqrt{3}M_{\rm Pl}}(t-t_0)\right]
\ee
Substituting it into eq.~(\ref{fried}) and using the definition $H=\dt{a}/a$
gives rise to a differential equation on the scale factor $a(t)$. Its solution 
is 
\be \lb{scalef}
a(t) = e^{H_0t} \left[ \fracmm{t-t_0}{\rm const.}\right]^{-3/4} 
\ee
where we have introduced the notation $H_0 =M/\sqrt{24}$. The presence of a 
singularity at $t=t_0$ in eq.~(\ref{scalef}) is harmless because our 
inflationary solution is only valid during the slow-roll inflation when 
$t\geq t_{\rm start}>t_0$, so that
it does not apply to the Big Bang. A resolution of the Big Bang singularity is 
supposed to require the higher-order curvature terms in the gravitational
effective action (\ref{action}).

The amount of inflation is measured by the e-foldings number
\be \lb{efol}
N_e = \int^{t_{\rm end}}_t H dt \approx 
\fracmm{1}{M^2_{\rm Pl}} \int^{\f}_{\f_{\rm end}} \fracmm{V}{V'} d\f
\ee
where the $t_{\rm end}$ stands for the (time) end of inflation when one of the
slow-roll parameters becomes equal to $1$. The number of e-foldings  between 
$50$ and $100$ is usually considered to be acceptable.

In the case of the slow-roll inflation with the scalar potential (\ref{srpot}),
we find that $\ve(\f)$ {\it first} approaches $1$ at $\f_{\rm end}=
\sqrt{\fracmm{3}{2}}M_{\rm Pl}\ln\left( 2\sqrt{3}-3\right)\approx
-0.94~M_{\rm Pl}$, since $\abs{\h(\f)}$ approaches $1$ later, at 
$\f_{\rm end}= -\sqrt{\fracmm{3}{2}}M_{\rm Pl}\ln\fracmm{5}{3}\approx
-0.62~M_{\rm Pl}$. Then eq.~(\ref{efol}) yields
\be \lb{efol1}
N_e =\fracmm{3}{4}\left( e^{-y} + y\right) -\fracmm{3}{4}\left(
\exp \left[ \sqrt{\fracmm{2}{3}}\cdot 0.94\right] -\sqrt{\fracmm{2}{3}} \cdot
0.94\right)\approx \fracmm{3}{4}\left( e^{-y} + y\right) -1.04
\ee
where we have used the notation (\ref{newvar}). Similarly, we find
\be \lb{ee2}
\ve = \fracmm{4e^{2y}}{3\left( 1-e^y \right)^2}
\qquad  {\rm and} \qquad 
\eta = \fracmm{-4e^y(1-2e^y)}{3\left( 1-e^y \right)^2}
\ee
Equation (\ref{efol1}) can now be used to get $y$ in terms of $N_e$, while
a substitution of $y(N_e)$ into eq.~(\ref{ee2}) yields both $\ve(N_e)$ and 
$\eta(N_e)$. The results of our numerical calculations  
(by using MATHEMATICA) are summarized in Table 1. 

An analytic approximation can be obtained by using the expansion with respect 
to the {\it inverse\/} number of e-foldings. For instance, eq.~(\ref{efol1}) 
yields
\be \lb{appr1}
 e^y = \fracmm{3}{4N_e} - \fracmm{9\ln N_e}{16N_e^2} -\fracmm{0.94}{N^2_e}
+{\cal O}\left( \fracmm{\ln^2 N_e}{N^3_e}\right) 
\ee
Equation (\ref{ee2}) now implies
\be \lb{apeps}
\ve = \fracmm{3}{4N^2_e} +{\cal O}\left( \fracmm{\ln^2 N_e}{N^3_e}\right) 
\ee
and
\be \lb{apeta}
\eta = - \fracmm{1}{N_e} + \fracmm{3\ln N_e}{4N_e^2} +\fracmm{5}{4N^2_e}
+{\cal O}\left( \fracmm{\ln^2 N_e}{N^3_e}\right) 
\ee

We are now ready for a calculation of the CMB {\it observable} quantitites in
our inflationary model, ie. for its specific physical predictions. The 
primordial spectrum in the power-law approximation takes the form of $k^{n-1}$
in terms of the comoving wave number $k$ and the spectral index $n$. In 
particular, the slope $n_s$ of the {\it scalar} power spectrum, associated
with the density perturbations, is given 
by \cite{llbook}  
\be \lb{sind}
n_s  = 1+2\h -6\ve~~, \ee
the slope of the {\it tensor} primordial spectrum, associated with the
gravitational waves, is given by \cite{llbook}
\be \lb{tind} n_t= -2\ve~~, \ee
whereas the scalar-to-tensor ratio is given by \cite{llbook} 
\be   r= 16\ve~~. \ee

Equations (\ref{apeps}), (\ref{apeta}) and (\ref{sind}) in our
model imply 
\be \lb{apns}
 n_s =1 - \fracmm{2}{N_e} + \fracmm{3\ln N_e}{2N_e^2} -\fracmm{2}{N^2_e}
+{\cal O}\left( \fracmm{\ln^2 N_e}{N^3_e}\right) 
\ee

The spectral indices are constrained by cosmological observations --- see eg., 
the recent WMAP5 data \cite{wmap5} that implies
\be \lb{sind5} 
 n_s = 0.960 \pm 0.013 \qquad {\rm and}\qquad r < 0.22 \ee
In addition, the amplitude of the initial perturbations, 
$\D^2_R=M^4_{\rm Pl}V/(24\p^2\ve)$,  is yet another physical observable, whose 
experimental value is \cite{llbook} 
\be \lb{ampl}
\left(\fracmm{V}{\ve}\right)^{1/4} =0.027\,M_{\rm Pl}=
6.6\times 10^{16}~{\rm GeV}
\ee
Equation (\ref{ampl}) determines the normalization of the $R^2$-term in 
eq.~(\ref{action}) as
\be \lb{scale} \fracmm{M}{M_{\rm Pl}}= 4\cdot \sqrt{\fracmm{2}{3}}\cdot (2.7)^2
\cdot \fracmm{e^y}{(1-e^y)^2} \cdot 10^{-4} = (3.5\pm 1.2)\cdot 10^{-5}
\ee
where, in the last step, we have used the value of $N_e=53.8\pm 18$, as it 
follows from eqs.~(\ref{apns}) and (\ref{sind5}). The results of our numerical 
calculations of the spectral indices are collected in  Table 1. In particular, 
we find that the WMAP5 experimental bounds on the scalar spectral index in 
eq.~(\ref{sind5}) are satisfied in the cosmological model (\ref{action}) 
provided that the e-foldings number $N_e$ lies between $35.9$ and $71.8$, with 
the middle value of $\bar{N}_e=53.8$. We also find the noticable suppression of
tensor fluctuations as $\abs{r}<8.2\cdot 10^{-3}$ and $\abs{n_t}<10^{-3}$. 
There is a possibility of further theoretical modification, which 
would imply more tuning of the spectral indices, when more terms of the 
higher-order in the curvature are added into the action (\ref{action}).

\begin{table}
\caption{The slow-roll parameters and spectral indices for some values 
of $N_{e}$}
\begin{center}
\begin{tabular}{c|cccc|c}
\hline
$N_e$  &  $\varepsilon (\times 10^{- 4})$  &  $\eta (\times 10^{- 2})$  &  
$r (\times 10^{-3})$  &  $n_{t} (\times 10^{- 4})$  &  $n_{s}$  \\
\hline
35  &  5.13  &  - 2.56  &  8.20  &  - 10.3  &  0.946  \\
40  &  3.99  &  - 2.27  &  6.39  &  - 7.98  &  0.952  \\
45  &  3.20  &  - 2.03  &  5.12  &  - 6.40  &  0.957  \\
50  &  2.62  &  - 1.84  &  4.19  &  - 5.24  &  0.962  \\
55  &  2.19  &  - 1.69  &  3.50  &  - 4.37  &  0.965  \\
60  &  1.85  &  - 1.55  &  2.96  &  - 3.71  &  0.968  \\
65  &  1.59  &  - 1.44  &  2.54  &  - 3.18  &  0.970  \\
70  &  1.38  &  - 1.34  &  2.21  &  - 2.76  &  0.972  \\
75  &  1.21  &  - 1.26  &  1.93  &  - 2.42  &  0.974  \\
\end{tabular}
\end{center}
\end{table}

\section{Conclusion}

Our main results are given by eqs.~(\ref{scalef}), (\ref{efol1}), (\ref{ee2}),
(\ref{apns}), (\ref{scale}) and Table 1. The leading terms agree with the known
results \cite{mchi,bar}. We confirm that the simplest (Starobinsky) model of 
$(R+R^2)$ gravity with the single new parameter $M$ may theoretically describe 
inflation and still agree with the experimental (CMB) observations. As regards 
the possible extensions to the quartic curvature terms, see eg., 
refs.~\cite{iihk}. The $f(R)$ gravity is extendable to $F(R)$ supergravity 
\cite{gket}.

\vglue.2in

\end{document}